\documentclass[12pt]{article}
\usepackage[margin=0.9in]{geometry}
\usepackage{import}
\usepackage{amsmath,amssymb}
\usepackage{mathtools}
\mathtoolsset{showonlyrefs,showmanualtags}
\usepackage{enumitem}
\usepackage{cancel}
\usepackage{graphicx}
\usepackage{bbm}
\usepackage[normalem]{ulem}
\usepackage{hyperref}
\usepackage{amsthm}
\usepackage[ruled]{algorithm2e}
\usepackage{multirow,array}

\usepackage{siunitx}

\usepackage{standalone}
\usepackage{pgfplots}
\usepackage{tikz}
\usepackage{subcaption}

\usepackage{import}
\usepackage{amsmath,amssymb}
\usepackage{mathtools}
\usepackage{etoolbox}

\makeatletter
\let\pragma@iinput=\@iinput
\def\@iinput#1{\xdef\@pragmafile{#1}\pragma@iinput{#1}}
\def\@pragmafile{default}
\def\pragmaonce{%
	\csname pragma@\@pragmafile\endcsname
	\global\expandafter\let \csname pragma@\@pragmafile\endcsname =  
}
\makeatother

\DeclareMathOperator*{\argmin}{arg\,min}

\DeclareMathOperator*{\sign}{sgn}

\ifdefined\ceil
\else
\DeclarePairedDelimiter\ceil{\lceil}{\rceil}

\fi
\DeclarePairedDelimiterX{\inprd}[2]{\langle}{\rangle}{#1, #2}

\newcommand{\Lone}[1]{\lVert #1 \rVert_1}
\newcommand{\Ltwo}[1]{\lVert #1 \rVert_2}

\newcommand{\EV}{{\mathbb{E}}}

\newcommand{\Rel}{\mathbb{R}}

\newcommand{\iid}{i.i.d.}

\newcommand{\cc}[1]{\mathcal{#1}}

\newcommand{\calF}{\cc{F}}

\definecolor{darkgreen}{rgb}{0.0, 0.5, 0.0}

\newcommand{\stkout}[1]{\ifmmode\text{\sout{\ensuremath{#1}}}\else\sout{#1}\fi}

\newcommand{\fig}[1]{{Fig.~\ref{fig:#1}}}
\newcommand{\tbl}[1]{{Table~\ref{tbl:#1}}}
\newcommand{\secn}[1]{{Sec.~\ref{secn:#1}}}

\newcommand{\eqn}[1]{{\eqref{eqn:#1}}}

\makeatletter
\newcommand*\ifcounter[1]{%
	\ifcsname c@#1\endcsname
	\expandafter\@firstoftwo
	\else
	\expandafter\@secondoftwo
	\fi
}
\makeatother

\ifcsmacro{theorem}{}{

	\ifcounter{chapter}{
		
	}{
		
	}
	
	\newcounter{example}[section]
	
	\newcounter{problem}[section]
	
}

\def\enumtheoremstart{\begin{enumerate}[noitemsep,label=(\roman*)]}
	\def\enumtheoremend{\end{enumerate}}

\usepackage{transparent}

\newif\ifshowcomments
\newif\ifshowdeleted

\newcommand{\devnull}[1]{}

\newcommand{\comment}[2][]{\ifshowcomments{\printcomment{#1}{#2}}\else\ignorespaces\fi}
\newcommand{\printcomment}[2]{\incolor{#1}{[[\ifx#1\empty\else#1: \fi#2]]}}

\newcommand{\ifequals}[3]{\ifthenelse{\equal{#1}{#2}}{#3}{}}
\newcommand{\case}[2]{#1 #2} 
\newenvironment{switch}[1]{\renewcommand{\case}{\ifequals{#1}}}{}

\definecolor{darkred}{rgb}{0.8, 0.01, 0.1}
\definecolor{darkgreen}{rgb}{0.0, 0.5, 0.0}
\definecolor{darkorange}{rgb}{0.93, 0.35, 0.1}

\newcommand{\incolor}[2]{\ignorespaces
	\begin{switch}{#1}\ignorespaces
		\case{TA}{\color{darkred}}\ignorespaces
		\case{SD}{\color{darkorange}}\ignorespaces
		\case{SK}{\color{darkgreen}}\ignorespaces
		\case{}{\color{red}}\ignorespaces
		#2
	\end{switch}
}

\title{
Rate distortion comparison of a few gradient quantizers
}

\author{Tharindu Adikari   \\
University of Toronto
}

\date{} 

\begin{document}

\maketitle

\begin{abstract}
This article is in the context of gradient compression. Gradient compression is a popular technique for mitigating the communication bottleneck observed when training large machine learning models in a distributed manner using gradient-based methods such as stochastic gradient descent. In this article, assuming a Gaussian distribution for the components in gradient, we find the rate distortion trade-off of gradient quantization schemes such as Scaled-sign and Top-$K$, and compare with the Shannon rate distortion limit. A similar comparison with vector quantizers also is presented.
\end{abstract}


\pragmaonce  

\newcommand{\vvw}{{w}}
\newcommand{\vvwu}[1]{\vvw_{#1}}
\newcommand{\vvwd}[1]{\vvw^{#1}}
\newcommand{\vvwud}[2]{\vvw^{#2}_{#1}}
\newcommand{\vvwji}{{\vvwud{j}{i}}}
\newcommand{\vvwki}{{\vvwud{k}{i}}}
\newcommand{\vvwkj}{{\vvwud{k}{j}}}

\newcommand{\gloss}{g}
\newcommand{\loss}{f}
\newcommand{\lossn}[1]{f_{\text{#1}}}
\newcommand{\lossSum}{{\frac{1}{2}(\loss_1(\vvw)+\loss_2(\vvw))}}

\newcommand{\sumwi}{{\sum}}
\newcommand{\sumworkersl}[1]{{\sum_{#1=1}^{\numDists}}}
\newcommand{\sumworkers}{{\sum_{i=1}^{\numDists}}}
\newcommand{\stepsize}{\eta}

\newcommand{\rvX}{{X}}
\newcommand{\rvXd}[1]{{\rvX_{#1}}}
\newcommand{\rvXi}{{\rvX_i}}
\newcommand{\rvXj}{{\rvX_{j}}}
\newcommand{\rvXij}{{\rvX^{i}_{j}}}

\newcommand{\distQ}{{Q}}
\newcommand{\distQi}{\distQ_{i}}
\newcommand{\numDists}{{n}}
\newcommand{\dimw}{{d}}
\newcommand{\batchSize}{{B}}
\newcommand{\dataSize}{{m}}

\newcommand{\wOptGD}{\vvwu{*}}
\newcommand{\wOptGDd}[1]{\vvwu{*}^{#1}}
\newcommand{\wOptSync}{\vvwu{\text{sync}}}

\newcommand{\Eft}{{\EV_{\rvX\sim \distQ}[\loss(\vvwu{t}, \rvX)]}}
\newcommand{\Ef}{{\EV_{\rvX\sim \distQ}[\loss(\vvw, \rvX)]}}
\newcommand{\EFunc}{F}
\newcommand{\gEFunc}{\nabla\EFunc}
\newcommand{\EF}{{\EFunc(\vvw)}}
\newcommand{\EFi}{{\EFunc_i(\vvw)}}
\newcommand{\EFj}{{\EFunc_j(\vvw)}}
\newcommand{\EFiNOw}{{\EFunc_i}}
\newcommand{\Efit}{{\EV_{\rvXi\sim \distQi}[\loss(\vvwu{t}, \rvXi)]}}
\newcommand{\Efi}{{\EV_{\rvX\sim \distQi}[\loss(\vvw, \rvX)]}}
\newcommand{\gEFw}{{\nabla\EF}}
\newcommand{\gEF}{{\gEFunc}}
\newcommand{\gEFiw}{{\nabla\EFi}}
\newcommand{\gEFjw}{{\nabla\EFj}}
\newcommand{\gEFi}{{\nabla\EFunc_i}}
\newcommand{\gEFj}{{\nabla\EFunc_j}}
\newcommand{\Egt}{\EV_{\rvX\sim \distQ}[\gloss(\vvwu{t}, \rvX)]}
\newcommand{\Eg}{\EV_{\rvX\sim \distQ}[\gloss(\vvw, \rvX)]}
\newcommand{\EG}{G(\vvw)}
\newcommand{\EGTheta}[1]{G(#1)}
\newcommand{\EGi}{G_i(\vvw)}
\newcommand{\EGj}{G_j(\vvw)}
\newcommand{\EGiTheta}[1]{G_i(#1)}
\newcommand{\Egn}[1]{{\EV_{\rvXd{#1}\sim \distQ_{#1}}[\gloss(\vvw, \rvXd{#1})]}}
\newcommand{\Egi}{\Egn{i}}
\newcommand{\Egit}{{\EV_{\rvXi\sim \distQi}[\gloss(\vvwu{t}, \rvXi)]}}
\newcommand{\EgiTheta}[1]{{\EV_{\rvXi\sim \distQi}[\gloss(#1, \rvXi)]}}
\newcommand{\EgiThetaji}{\EgiTheta{\vvwji}}

\newcommand{\iteratescriptstyle}[1]{{#1}}
\newcommand{\momentum}{v}
\newcommand{\wi}[2][]{\vvw^{#1}_{#2}}
\newcommand{\twi}[2][]{\tilde{\vvw}^{#1}_{#2}}
\newcommand{\dgi}[2][]{\dot{\gloss}^{#1}_{#2}}
\newcommand{\gi}[2][]{\gloss^{#1}_{#2}}
\newcommand{\vi}[2][]{\momentum^{#1}_{#2}}
\newcommand{\hvi}[2][]{\hat{\momentum}^{#1}_{#2}}
\newcommand{\ai}[2][]{\gloss^{#1}_{#2}}
\newcommand{\bgi}[2][]{\tilde{\gloss}^{#1}_{#2}}
\newcommand{\acgi}[1]{\bar{v}_{#1}}
\newcommand{\ccgi}[1]{\fQ(\acgi{#1})}
\newcommand{\agi}[1]{\bar{\gloss}_{#1}}
\newcommand{\hgi}[2][]{\hat{\gloss}^{#1}_{#2}}
\newcommand{\tgi}[2][]{\fQ({\gloss}^{#1}_{#2})}
\newcommand{\ttgi}[2][]{\fQQ({\gloss}^{#1}_{#2})}
\newcommand{\Fi}[2][]{\calF^{#1}_{#2}}

\newcommand{\fQQ}{\fQ_1}
\newcommand{\fQ}{Q}

\newcommand{\eQ}{s}
\newcommand{\compU}{U}
\newcommand{\sigQ}{\sigma_\compU}
\newcommand{\sigQQ}{\sigma_{\fQQ}}

\newcommand{\avgopn}{\frac{1}{\numDists}\sum_{i\in[\numDists]}}
\newcommand{\compress}{\text{\normalfont{\textit{encode}}}}
\newcommand{\decompress}{\text{\normalfont{\textit{decode}}}}
\newcommand{\datai}[2][]{\mathcal{G}^{#1}_{#2}}
\newcommand{\ei}[2][]{e^{#1}_{#2}}
\newcommand{\ri}[2][]{r^{#1}_{#2}}
\newcommand{\gti}[2][]{\tilde{\nabla}^{#1}_{#2}}
\newcommand{\gdi}[2][]{\nabla^{#1}_{#2}}
\newcommand{\ghi}[2][]{\hat{g}^{#1}_{#2}}
\newcommand{\gbi}[2][]{\bar{g}^{#1}_{#2}}
\newcommand{\ppi}[2][]{u^{#1}_{#2}}
\newcommand{\tri}[2][]{\fQ(\ri[#1]{#2})}

\newcommand{\Ei}[2][]{E^{#1}_{#2}}
\newcommand{\Gi}[2][]{{G}^{#1}_{#2}}
\newcommand{\Gdi}[2][]{\dot{G}^{#1}_{#2}}
\newcommand{\Ghi}[2][]{\hat{G}^{#1}_{#2}}
\newcommand{\bp}[1]{\bar{p}_{#1}}

\newcommand{\numlocsteps}{c}
\newcommand{\binentropy}{{H}}
\newcommand{\probone}{{p}}
\newcommand{\vecb}{{b}}
\newcommand{\encin}{{x}}
\newcommand{\encref}{{y}}

\newcommand{\totbits}{{B_\text{\normalfont X}}}
\newcommand{\totbitsS}{{B_\text{\normalfont S}}}
\newcommand{\bitratio}{{\totbits/\totbitsS}}

\newcommand{\fQy}{{\fQ_{\alpha}^{\encref}}}
\newcommand{\fQyo}{{\fQ_{1}^{\encref}}}
\newcommand{\fQyx}{{\fQy(\encin)}}

\newcommand{\upperbndalpha}{\frac{1}{4}\left(1-\sqrt{1-\frac{1}{\dimw}}\right)^2}

\newcommand{\setbin}{\{0,1\}}
\newcommand{\setbind}{\setbin^\dimw}
\newcommand{\setA}{\{+a,-a\}}
\newcommand{\setAd}{\setA^\dimw}

\newcommand{\Xd}{X^\dimw}
\newcommand{\hXd}{\hat{X}^\dimw}
\newcommand{\hX}{\hat{X}}
\newcommand{\xd}{x^\dimw}
\newcommand{\hu}{\hat{u}}

\section{Introduction} \label{secn:bacground}

Rate distortion theory provides a mathematical framework for analyzing the performance of lossy compression schemes. 
In work related to machine learning, rate distortion theory has been used to analyze model compression \cite{gao2019rate}. 
In order to fit gradient compression within the rate distortion framework we assume a Gaussian distribution to describe the distribution of the components in the gradient vector. 
The empirical results provided in \cite{glorot2010understanding, bernstein2018signsgd, shi2019understanding} justifies this assumption. 
Specifically, the authors of \cite{glorot2010understanding} study the distribution of gradient components when training feed forward deep neural networks on datasets such as MNIST, CIFAR-10 and Small-ImageNet. 
They empirically show (see Figure~7 therein) that the gradient components have Gaussian-like distributions. 
This observation is corroborated by the experimental results presented in \cite{bernstein2018signsgd} and \cite{shi2019understanding}. 
The authors of \cite{bernstein2018signsgd} empirically study the distribution of components in the stochastic gradient for convolutional neural network models of practical interest such as ResNet-20 and ResNet-50, trained on CIFAR-10 and Imagenet datasets respectively. 
The authors demonstrate (see Remark 1 therein) that the distribution of components in a gradient vector is uni-modal, symmetric and can be approximated by a zero-centered Gaussian distribution. 
A similar study is carried out in \cite{shi2019understanding} where the results confirm (see Appendix A.2 therein) the uni-modality and symmetricity of the gradient distribution. 
The authors also demonstrate that the statistics of elements in the gradient may change over the epochs (see Figure~2 and 8 in \cite{shi2019understanding}). 
These observations provide grounds for assuming a Gaussian distribution to describe the distribution of gradient components.

Let $X\in\Rel$ with $X\sim\mathcal{N}(0,\,\sigma^{2})$ be a symbol passed to a quantizer and let $\hX\in\Rel$ be the quantizer output, i.e., the reconstruction of $X$. The reconstruction is a discrete random variable and let it be represented with a rate $R$. 
Let $D$ be the distortion achieved with a rate $R$. 
The Shannon rate distortion function $R(D)$ provides the theoretical lower bound for the achievable rate $R$ for a distortion $D$. 
If squared error is taken as the distortion measure the rate distortion function for a Gaussian source is $R(D) = \frac{1}{2}\log\frac{\sigma^2}{D}$ for $0\leq D\leq\sigma^2$ and $R(D)=0$ otherwise \cite[p.~344]{cover1999elements}. 
The logarithm is to the base $2$. 
This rate can only be achieved by considering block descriptions, i.e., a large number of realizations of $X$ compressed together. 
In gradient compression one can compress together a large number of components in the gradient vector. 
The rate distortion lower bound presented above only applies when the compressed elements are independent. 
Although this may not be the case for the components in the gradient vector, $R(D)$ is a useful theoretical bound to compare the performance of existing gradient quantizers. 
In our analysis we assert squared norm of the compression error as the distortion measure as it is has been widely used in gradient compression literature \cite{stich2018sparsified, karimireddy2019error}.

\begin{figure}
\centering\includegraphics[width=0.97\textwidth]{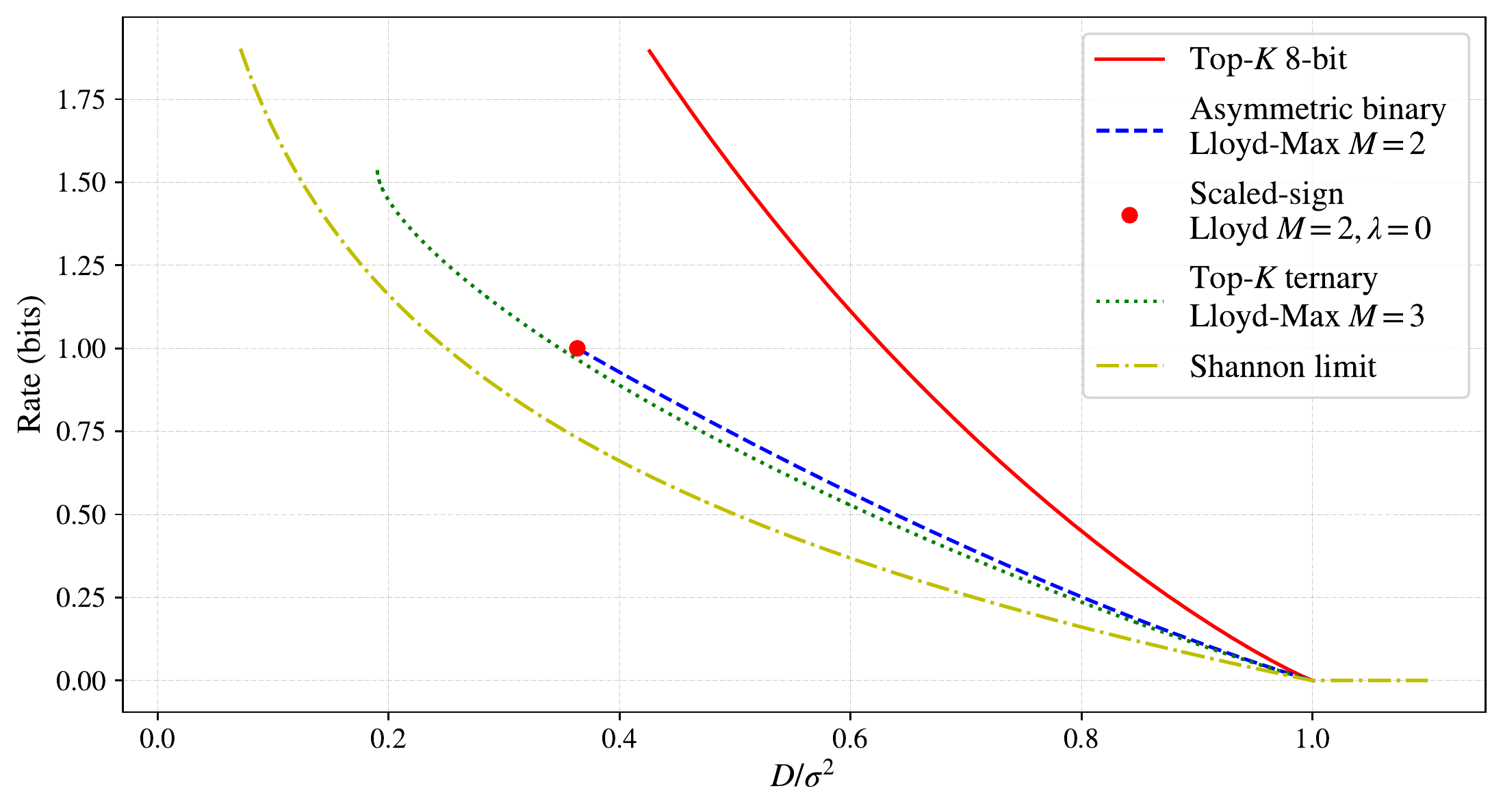}
\caption{Comparing the rate distortion trade-off of different quantizers with the Shannon limit. 
To see the derivation of the plots refer to \secn{derivations}. 
Specifically, refer to 
\secn{scaledsign} for Scaled-sign, 
\secn{asymmetricbin} for Asymmetric binary, 
\secn{topkeightbit} for Top-$K$ 8-bit,
and
\secn{topkternary} for Top-$K$ ternary.
} 
\label{fig:rate_dist_gauss}
\end{figure}

\begin{figure}
\centering\includegraphics[width=0.97\textwidth]{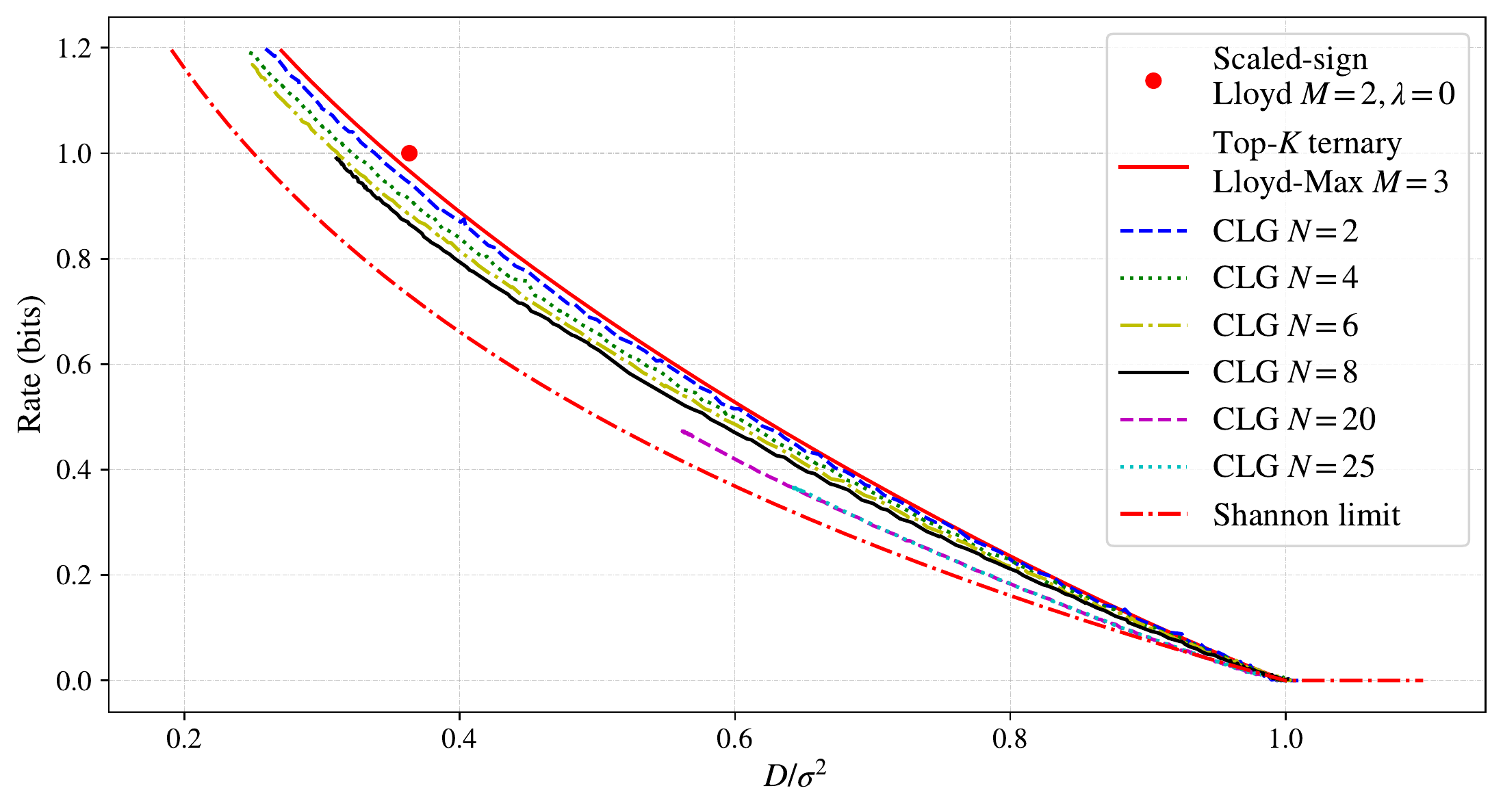}
\caption{Additional comparisons with vector quantizers with vector dimension $N$. As $N$ gets large vector quantizers exhibit a clear advantage over the scalar quantizers, albeit at a high computational cost. The details of vector quantizers are presented in \secn{ecvq}.} 
\label{fig:rate_dist_clg}
\end{figure}

In this article we find the rate distortion trade-off of compression schemes such as Scaled-sign \cite{bernstein2018signsgd} and Top-$K$ \cite{dryden2016communication}, and compare their performances with the Shannon bound $R(D)$. 
In \secn{ecsq} we start with setting up the scalar quantization problem. 
In \secn{derivations} we present the derivations of the rate distortion trade-off of the scalar quantizers considered. 
Our comparison is presented in \fig{rate_dist_gauss}. The substantial gap to the Shannon limit can be closed with vector quantization. 
\fig{rate_dist_clg} presents a similar comparison to that of \fig{rate_dist_gauss}, but with vector quantizers. 
The details of the vector quantizers is presented in \secn{ecvq}. 
However, it must be noted that the vector quantization may not be a feasible solution for gradient compression due to the large computation requirements involved. 

\section{Optimal scalar quantizers} \label{secn:ecsq}
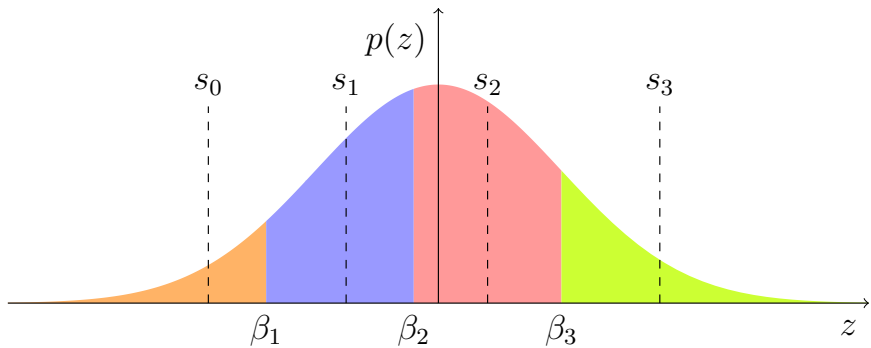
\begin{figure}
\centering
\pgfmathdeclarefunction{gauss}{2}{%
\pgfmathparse{1/(#2*sqrt(2*pi))*exp(-((x-#1)^2)/(2*#2^2))}%
}

\begin{tikzpicture}
\def\scl{4}
\def\sig{1.5}
\def\normaltwo{\x,{\scl*exp(-0.5*(\x/\sig)^2)/\sig}}
\def\hgt{0.6*\scl}

\def\xmin{-3.5*\sig}
\def\szro{-1.87*\sig}
\def\tone{-1.4*\sig}
\def\sone{-0.75*\sig}
\def\ttwo{-0.2*\sig}
\def\stwo{0.4*\sig}
\def\tthr{1.0*\sig}
\def\sthr{1.8*\sig}
\def\xmax{-\xmin}

\fill [fill=orange!60] ({\xmin},0) -- plot[domain={\xmin}:{\tone},samples=100] (\normaltwo) -- ({\tone},0) -- cycle;
\fill [fill=blue!40] ({\tone},0) -- plot[domain={\tone}:{\ttwo},samples=100] (\normaltwo) -- ({\ttwo},0) -- cycle;
\fill [fill=red!40] ({\ttwo},0) -- plot[domain={\ttwo}:{\tthr},samples=100] (\normaltwo) -- ({\tthr},0) -- cycle;
\fill [fill=lime!80] ({\tthr},{0}) -- plot[domain={\tthr}:{\xmax}] (\normaltwo) -- ({\xmax},0) -- cycle;

\draw[] ({\tone},0) node[below] {$\beta_1$};
\draw[] ({\ttwo},0) node[below] {$\beta_2$};
\draw[] ({\tthr},0) node[below] {$\beta_3$};
\draw[dashed] (\szro,0) -- (\szro,\hgt) node[above] {$s_0$};
\draw[dashed] (\sone,0) -- (\sone,\hgt) node[above] {$s_1$};
\draw[dashed] (\stwo,0) -- (\stwo,\hgt) node[above] {$s_2$};
\draw[dashed] (\sthr,0) -- (\sthr,\hgt) node[above] {$s_3$};

\draw[->] (0,0) -- (0,\scl*0.9) node[left] {};
\draw (0,\scl*0.8) node[left] {$p(z)$};

\draw[->] (\xmin,0) -- (\xmax,0) node[right] {};
\draw (\xmax,-0.3) node[left] {$z$};

\end{tikzpicture}
\caption{A scalar quantizer for $M=4$. The PDF of the source is indicated as $p(z)$.}
\label{fig:quantizeRegionsK}
\end{figure}

In scalar quantization the realizations of $X$ are quantized separately, i.e., without block descriptions. 
Since we are considering a single scalar, we can use the real line to illustrate the idea. 
The real line is separated into $M$ quantization regions, and all source values corresponding one region are assigned the same reconstruction point. 
\fig{quantizeRegionsK} illustrates one possible assignment when $M=4$. 
The quantization regions are separated by thresholds $\beta_1, \beta_2$ and $\beta_3$, and the reconstruction of $x$ values in the four regions from left to right are given by $s_0, s_1, s_2$ and $s_3$. 
For instance, given that $\beta_1\leq X\leq \beta_2$ the reconstruction is $\hX=s_1$. 

\subsection{ECSQ problem} 
Let $p_0,\dots,p_{M-1}$ be the probability masses of the $M$ regions 
corresponding to $s_0,\dots,s_{M-1}$. 
The \emph{entropy-constrained scalar quantization problem} (ECSQ) \cite{wiegand2011source} is to solve for the threshold and reconstruction points that minimize the expected distortion, subject to a maximum entropy constraint. 
The entropy-constrained scalar quantization problem is expressed as 
\begin{align}
\min_{\begin{subarray}{l}
s_0,\dots,s_{M-1}\\
\beta_1,\dots,\beta_{M-1}
\end{subarray}} \quad & \EV[(X-\hX)^2]\\
\textrm{s.t.} \quad\quad & H(\hX) \leq R, 
\end{align}
where $H(\hX) = - \sum_{m=0}^{M-1} p_m\log p_m$. 
The problem is converted to an unconstrained optimization problem by writing the Lagrangian and minimizing 
\begin{align}
J = \EV[(X-\hX)^2] + \lambda(H(\hX)-R). \label{eqn:lagrangian}
\end{align}
Note that for $\lambda=0$ the optimization problem reduces to minimization of distortion alone, without the entropy constraint. 
As one may expect, the minimal distortion in this case is obtained when the entropy is at a maximum. 
This happens when all quantization regions are equiprobable, i.e. when $p_m=1/M$.

\subsection{Lloyd-Max algorithm} 
The solution to minimization of $J$ is obtained by the application of the iterative Lloyd-Max algorithm \cite{lloyd1982least}. 
The Lloyd-Max algorithm is executed as follows. Let $p(x)$ be the PDF of the source $X$. 
First, the reconstruction points $s_0,\dots,s_{M-1}$ are randomly initialized. Second, the thresholds and reconstruction points are iteratively updated as 
\begin{align}
\beta_m &= \frac{s_{m-1}+s_m}{2} - \lambda\frac{\log p_{m-1}-\log p_{m}}{2(s_{m-1}-s_m)}
\qquad\qquad\qquad \,\,\, \text{ for } m=1,\dots,M-1 \\
p_m &= \int_{\beta_m}^{\beta_{m+1}} p(x)\,dx, \qquad
s_m = \frac{\int_{\beta_m}^{\beta_{m+1}} xp(x)\,dx}{\int_{\beta_m}^{\beta_{m+1}} p(x)\,dx} 
\qquad\quad \text{ for } m=0,\dots,M-1, \label{eqn:reconptlloyd}
\end{align}
where we define $\beta_0=-\infty$ and $\beta_M=\infty$. 
The iteration is continued until no further significant reduction in $J$ is observed. 
For different $\lambda\geq0$ we obtain operating points of the quantizer corresponding to different entropy rates. 
Note that as per \eqn{reconptlloyd}, for a given quantization region $(\beta_m, \beta_{m+1})$, the optimal reconstruction point $s_m$ is the centroid of the region. 
This can be verified by observing that for any random variable $Y$ and a constant $a$, $\EV[(Y-a)^2]$ is minimized by setting $a=\EV[Y]$.

\section{Derivations of rate distortion trade-off} \label{secn:derivations}
\subsection{Scaled-sign: $M=2$, $\lambda=0$} \label{secn:scaledsign}
The Scaled-sign quantization \cite{bernstein2018signsgd, seide20141} can be recovered by setting $M=2$ and $\lambda=0$ in \eqn{lagrangian}. 
Scaled-sign, i.e., setting $\hX$ a scaled version of the sign of $X$, is optimal when the rate $R=1$ and block descriptions are not allowed. 
The optimum reconstruction points are given by the centroids of the two quantization regions. 
This can be verified by using \eqn{reconptlloyd}. For $M=2$ and $\lambda=0$ we have 
\[
s_1
= \frac{\int_{0}^{\infty} xp(x)\,dx}{\int_{0}^{\infty} p(x)\,dx} 
= 2\int_{0}^{\infty} xp(x)\,dx 
= \int_{-\infty}^{0} -xp(x)\,dx + \int_{0}^{\infty} xp(x)\,dx, 
\]
which gives 
\[
s_1
= \int_{-\infty}^{0} |x|p(x)\,dx + \int_{0}^{\infty} |x|p(x)\,dx 
= \int_{-\infty}^{\infty} |x|p(x)\,dx 
= \EV[|X|]. 
\]
When $p(x)$ is Gaussian $\EV[|X|]=\sigma\sqrt{2/\pi}$. 
Therefore, the relationship between $X$ and $\hX$ can be compactly expressed as $\hX=\sigma\sqrt{2/\pi} \sign(X)$. 
If we have $d$ observations of the random variable $X$ represented as components in a vector $u\in\Rel^\dimw$, we can approximate $\EV[|X|]$ by computing $\frac{\Lone{u}}{\dimw}$. 
With this approximation we have $\hu = \frac{\Lone{u}}{\dimw}\sign(u)$. 
The encoded $\hu$ vector first stores $\frac{\Lone{u}}{\dimw}$ with finite precision, and second, stores $\sign(u)$ as a binary vector. 
Since the median of $p(x)$ is at zero, we expect the number of ones and zeros in the resulting binary vector to be equal. 
This means that the encoding of the binary vector takes one bit per component, and the vector cannot be compressed further without information loss. 
The comparison of Scaled-sign with other quantizers is presented in \fig{rate_dist_gauss}. 

The expected distortion $\EV[(X-\hX)^2]$ with Scaled-sign is $\frac{\pi-2}{\pi}\sigma^2 = 0.3633\sigma^2$. 
The distortion rate function for Gaussian $p(x)$ is $D(R)=\sigma^22^{-2R}$, which for $R=1$ is $0.25\sigma^2$. 
We have $0.3633\sigma^2$ greater than the optimal $0.25\sigma^2$ since scalar quantization does not allow block descriptions.

\subsection{Asymmetric binary quantizer: $M=2$, $\lambda\geq0$} \label{secn:asymmetricbin}
Scaled-sign provides a rate of $1$ bit with a fixed distortion. 
The Scaled-sign quantizer can be generalized to obtain all rates less than $1$ by setting $M=2$ and a non-zero $\lambda$ in \eqn{lagrangian}. 
For the ease of use let us refer to this generalization as the asymmetric binary quantizer. 
Instead of solving \eqn{lagrangian} using the Lloyd-Max algorithm, we can use the following method to obtain the rate distortion function for the asymmetric binary quantizer. 

The idea behind the asymmetric binary quantizer is to change the boundary that separates quantization regions in a way that the two regions have asymmetric probability masses. 
This causes the two reconstruction points to be selected with different probabilities. 
Let the new boundary be $\sigma\beta$ for some $\beta\in\Rel$, and let $\phi$ and $\Phi$ be the PDF and CDF of the standard normal distribution. 
\begin{figure}
	\centering
\pgfmathdeclarefunction{gauss}{2}{%
\pgfmathparse{1/(#2*sqrt(2*pi))*exp(-((x-#1)^2)/(2*#2^2))}%
}

\begin{tikzpicture}
\def\scl{4}
\def\sig{1.5}
\def\normaltwo{\x,{\scl*exp(-0.5*(\x/\sig)^2)/\sig}}

\def\cm{-0.4*\sig}
\def\cp{1.4*\sig}
\def\xmax{3.5*\sig}
\def\xmin{-\xmax}
\def\bt{0.6*\sig}

\def\fbt{\scl*exp(-0.5*(\bt/\sig)^2)/\sig}
\def\fcm{\scl*exp(-0.5*(\cm/\sig)^2)/\sig}
\def\fcp{\scl*exp(-0.5*(\cp/\sig)^2)/\sig}

\fill [fill=orange!60] ({\xmin},0) -- plot[domain={\xmin}:{\bt},samples=100] (\normaltwo) -- ({\bt},0) -- cycle;
\fill [fill=blue!40] ({\bt},{0}) -- plot[domain={\bt}:{\xmax}] (\normaltwo) -- ({\xmax},0) -- cycle;


\draw[] ({\bt},0) node[below] {$\sigma\beta$};
\draw[dashed] ({\cm},{\fcm}) -- ({\cm},0) node[below] {$s_0$};
\draw[dashed] ({\cp},{\fcp}) -- ({\cp},0) node[below] {$s_1$};

\draw[->] (0,0) -- (0,\scl*0.9) node[left] {};
\draw (0,\scl*0.8) node[left] {$p(x)$};

\draw[->] (\xmin,0) -- (\xmax,0) node[right] {};
\draw (\xmax,-0.3) node[left] {$x$};

\end{tikzpicture}
	\caption{Quantizing regions of the asymmetric binary quantizer. 
		All values to the left of $\sigma\beta$ and to the right of $\sigma\beta$ get mapped to $s_0$ and $s_1$ respectively. We recover Scaled-sign by setting $\beta=0$.} 
	\label{fig:quantizeRegionsASign}
\end{figure}
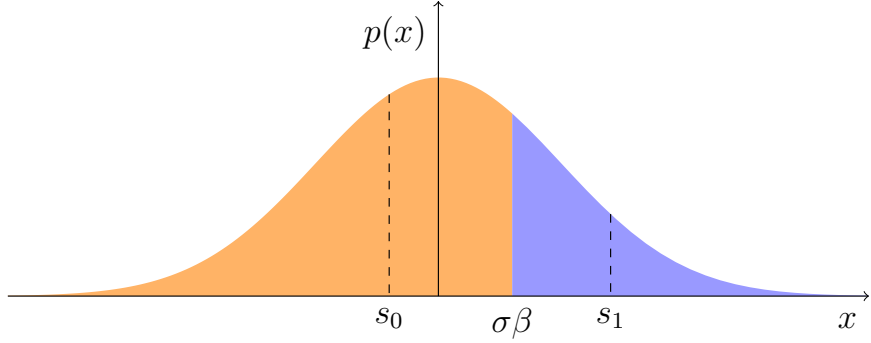
\fig{quantizeRegionsASign} illustrates the quantizing regions and the corresponding reconstruction points. 
We map $X<\sigma\beta$ to $s_0$ and $X>\sigma\beta$ to $s_1$ where $s_0$ and $s_1$ are yet to be determined. 
This means that we assign a random observation $X$ to $s_0$ with probability $\Phi(\beta)$ and to $s_1$ with probability $1-\Phi(\beta)$. 
These assignments can be represented by a binary vector which has entropy $H(\Phi(\beta))$. 
By changing $\beta$ we can attain any entropy between $1$ and $0$, which translates to the rate of the asymmetric binary quantizer. 
Expected distortion in this scheme is 
\begin{align}
\EV[(X-\hX)^2] 
&= \EV[(X-\hX)^2|X<\sigma\beta]\Pr(X<\sigma\beta) 
+ \EV[(X-\hX)^2|X>\sigma\beta]\Pr(X>\sigma\beta) \\
&= \EV[(X-s_0)^2|X<\sigma\beta]\Phi(\beta)
+ \EV[(X-s_1)^2|X>\sigma\beta](1-\Phi(\beta)). \label{eqn:expdistortiontotal}
\end{align}

For a given $\beta$ we can determine $s_0$ and $s_1$ that minimize the expected distortion. 
Note that for any random variable $Y$ and a constant $a$, $\EV[(Y-a)^2]$ is minimized by setting $a=\EV[Y]$. 
Therefore, we set $s_0$ to be the expected value of the normal distribution truncated from above at $\sigma\beta$. 
With this choice of $s_0$ the first term $\EV[(X-s_0)^2|X<\sigma\beta]$ in \eqn{expdistortiontotal} becomes the variance of the normal distribution truncated from above at $\sigma\beta$. 
A similar argument is applied to finding the $s_1$ and the resulting second term in \eqn{expdistortiontotal}. 
Plugging in the truncation limits to the variance expression of the truncated normal distribution and simplifying gives 
$\EV[(X-\hX)^2] = \sigma^2(h(\beta)+h(-\beta))$ where 
\[
h(\beta) = 
\left(1-\beta\frac{\phi(\beta)}{\Phi(\beta)}
-\left(\frac{\phi(\beta)}{\Phi(\beta)}\right)^2\right)\Phi(\beta). 
\]

So far we have established that for a given $\beta$ the asymmetric binary quantizer yields a rate $H(\Phi(\beta))$ and a distortion $\sigma^2(h(\beta)+h(-\beta))$. 
For a given distortion budget $D$, we can find the rate offered by this scheme by solving the optimization problem 
\begin{align}
\min_{\beta} \quad & H(\Phi(\beta))\\
\textrm{s.t.} \quad & \sigma^2(h(\beta)+h(-\beta))\leq D.
\end{align}
First, we observe that minimizing $H(\Phi(\beta))$ is equivalent to either minimizing or maximizing $\Phi(\beta)$. 
Since $\Phi$ is a monotonically increasing in $\beta$, we can equivalently minimize or maximize $\beta$ itself. 
Since both the objective function and the constraint are symmetric around $\beta=0$, without loss of generality we choose to maximize $\beta$ and cast the optimization problem as 
$\max_{\beta>0} \beta$ subject to $\sigma^2(h(\beta)+h(-\beta))\leq D$. 
One can plot the constraint against $\beta$ and verify that $h(\beta)+h(-\beta)$ is monotonically increasing for $\beta>0$, and that there is a one-to-one mapping between $\beta$ and $D$. 
This means that we can first find $\beta_D$ that achieves equality $\sigma^2(h(\beta_D)+h(-\beta_D))=D$, and then compute the rate as $H(\Phi(\beta_D))$. 
See \fig{rate_dist_gauss} to compare the rate distortion trade-off of the asymmetric binary quantizer with other quantizers. 

\subsection{Top-$K$ with $b$-bit precision: $M>3, b>1$} \label{secn:topkeightbit}

Sparsification schemes such as Top-$K$ preserves only the most significant elements in the gradient and achieves nearly consistent convergence performance with vanilla SGD. 
A subset of gradient entries with largest absolute values are selected and the rest is set to zero, essentially sparsifying the reconstructed gradient along the standard basis. 
The encoded representation comprises of the list of non-zero locations and the list of corresponding non-zero values. 
Sparsification can be achieved by either setting a threshold and zeroing out all elements whose magnitudes are smaller than the threshold, or by selecting a fixed size subset of gradient entries that have the largest magnitude. 
In \tbl{topklsummary} we summarize the Top-$K$ variants and indicate how in those variants the largest elements are selected, 
and how the reconstruction values are assigned. 
Thresholding means values smaller than a threshold are set zero. Largest $K$ means that all except the $K$ gradient entries with largest magnitude are set to zero. The reconstruction point is either the original value it self, or the average of the non-zero values, or the threshold. 

\begin{table}[]
\centering
\begin{tabular}{|c|c|c|}
\hline
\textbf{Reference}             		& \textbf{Sparsification method} & \textbf{Reconstruction point}    \\
\hline\hline
\cite{strom2015scalable}       		& Thresholding                   & Threshold (ternary)              \\\hline
\cite{dryden2016communication} 		& Thresholding                   & Average (ternary) 				\\\hline
\cite{aji2017sparse,lin2017deep}    & Thresholding       			 & Original ($b$-bit)    			\\\hline
\cite{stich2018sparsified,wangni2018gradient}	   & Largest $K$     & Original ($b$-bit)				\\\hline
\end{tabular}
\caption{Summarizing differences in sparsification methods.}
\label{tbl:topklsummary}
\end{table}

Consider the last two schemes in \tbl{topklsummary} that corresponding to $b$-bit reconstruction points. 
Assume that we have $d$ observations of the random variable $X$ represented as components in a vector $u\in\Rel^\dimw$. 
Similar to the approaches in \cite{aji2017sparse,lin2017deep}, for $\beta\geq0$, all values in $u$ that are in $(-\sigma\beta, \sigma\beta)$ interval are set to zero. 
The rest of the values are represented with $b$-bit precision. 
The non-zero locations can be represented by a binary vector. 
In expectation we have $2\Phi(-\beta)\dimw$ non-zero values which is similar to the $K$ in Top-$K$ in \cite{stich2018sparsified,wangni2018gradient}. 
For a given $\beta$, the expected rate of this scheme is $R = H(2\Phi(-\beta)) + 2\Phi(-\beta)b$. 
The first term is due to the entropy of the \iid{} binary vector that represents non-zero locations, 
and the second term is due to the $b$-bit representation of non-zero values. 
Both terms in $R$ are monotonically decreasing in $\beta$. 

We assume that $b$ is large enough so that the distortion incurred by encoding a real value with $b$-bit precision is negligible. 
In practice one often use $32$-bit single-precision floats for this task. 
Therefore, only the zero reconstruction values contribute to the distortion. 
The expected distortion can be computed using the expression for the variance of the normal distribution truncated from below and above at
$-\beta$ and $\beta$. 
This gives 
\[
\EV[(X-\hX)^2] = 
\sigma^2\left(1-\frac{2\beta\phi(\beta)}{2\Phi(\beta)-1}\right)(2\Phi(\beta)-1), 
\]
which, one can show is monotonically increasing in $\beta$. 
Therefore, for a given distortion budget $D$ we can find the $\beta$ that achieves equality $\EV[(X-\hX)^2]=D$ and find the corresponding rate. 
In \fig{rate_dist_gauss} we plot $R$ against $D/\sigma^2$ for $b=8$ and compare this Top-$K$ variant discussed in \cite{aji2017sparse,lin2017deep} with other quantizers. 
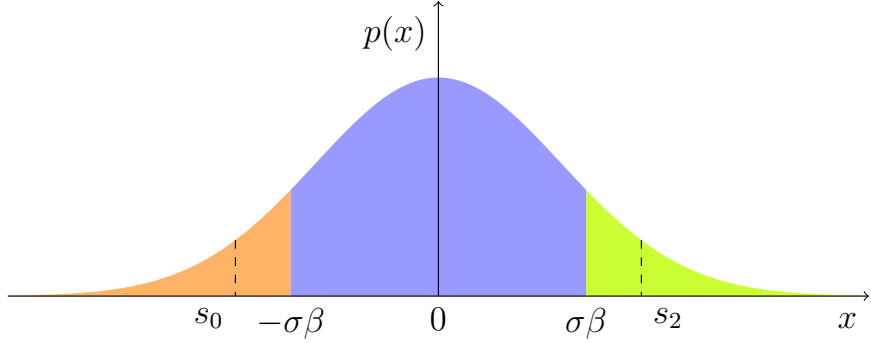
\begin{figure}
\centering
\pgfmathdeclarefunction{gauss}{2}{%
\pgfmathparse{1/(#2*sqrt(2*pi))*exp(-((x-#1)^2)/(2*#2^2))}%
}

\begin{tikzpicture}
\def\scl{4}
\def\sig{1.5}
\def\normaltwo{\x,{\scl*exp(-0.5*(\x/\sig)^2)/\sig}}

\def\cp{1.65*\sig}
\def\xmax{3.5*\sig}
\def\xmin{-\xmax}
\def\bt{1.2*\sig}

\def\fbt{\scl*exp(-0.5*(\bt/\sig)^2)/\sig}
\def\fcm{\scl*exp(-0.5*(-\cp/\sig)^2)/\sig}
\def\fcp{\scl*exp(-0.5*(\cp/\sig)^2)/\sig}

\fill [fill=orange!60] ({\xmin},0) -- plot[domain={\xmin}:{-\bt},samples=100] (\normaltwo) -- ({-\bt},0) -- cycle;
\fill [fill=blue!40] ({-\bt},0) -- plot[domain={-\bt}:{\bt},samples=100] (\normaltwo) -- ({\bt},0) -- cycle;
\fill [fill=lime!80] ({\bt},{0}) -- plot[domain={\bt}:{\xmax}] (\normaltwo) -- ({\xmax},0) -- cycle;


\draw[] ({\bt},0) node[below] {$\sigma\beta$};
\draw[] ({-\bt},0) node[below] {$-\sigma\beta$};
\draw[] (0,0) node[below] {$0$};
\draw[dashed] ({-\cp},{\fcm}) -- ({-\cp},0) node[below left] {$s_0$};
\draw[dashed] ({\cp},{\fcp}) -- ({\cp},0) node[below right] {$s_2$};

\draw[->] (0,0) -- (0,\scl*0.9) node[left] {};
\draw (0,\scl*0.8) node[left] {$p(x)$};

\draw[->] (\xmin,0) -- (\xmax,0) node[right] {};
\draw (\xmax,-0.3) node[left] {$x$};

\end{tikzpicture}
\caption{Quantizing regions of a sparsification-based quantizer. 
The reconstruction points corresponding to the three regions are $s_1, 0$ and $s_3$.}
\label{fig:quantizeRegionsTopk}
\end{figure}

\subsection{Top-$K$ with ternary alphabet: $M=3$, $\lambda\geq0$} \label{secn:topkternary}
Let us now consider the case when $b$ is small enough to cause non-negligible distortion. 
In particular, let us consider the case when all non-zero values are assigned one of two reconstruction values. The idea is illustrated in \fig{quantizeRegionsTopk}. 
This quantizer has three regions, and three reconstruction points. 
The rate distortion function of this scheme can be obtained by solving \eqn{lagrangian} with $M=3$ for different $\lambda$. 
For a given $\beta$, the optimal reconstruction points in \fig{quantizeRegionsTopk} are the centroids of the quantizing regions. 
This is equivalent to the scheme proposed in \cite{dryden2016communication}. 

Instead of solving \eqn{lagrangian} we can also find the rate distortion function of this scheme as follows. 
For a fixed $\beta$, from left to right, values belonging to three coloured regions get mapped to $s_0$, $0$ and $s_2$. 
Setting $\beta$ such that $s_0$ and $s_2$ become the centroids of the respective regions minimize the distortion. 
Letting $p_1 = \Phi(-\beta)$ and $p_0 = 1-2p_1$, we compute the expected distortion 
\begin{align}
\EV[(X-\hX)^2] 
&= 2\EV[(X-\hX)^2|X<-\sigma\beta]\Pr(X<-\sigma\beta) \\
&\quad + \EV[(X-\hX)^2||X|\leq\sigma\beta]\Pr(|X|\leq\sigma\beta) \\
&= 2\sigma^2\left(1+\frac{\beta\phi(\beta)}{p_1} - \left(\frac{\phi(\beta)}{p_1}\right)^2\right)p_1 
+ \sigma^2\left(1-\frac{2\beta\phi(\beta)}{p_0}\right)p_0. 
\end{align}
The expected rate is given by the entropy of the ternary vector that encodes three possible quantization values $s_1$, $0$ and $s_3$. 
Therefore, the rate is $R = -2p_1\log p_1 - p_0\log p_0$ in bits. 
We let $\EV[(X-\hX)^2]=D$ and plot $R$ versus $D/\sigma^2$ by varying $\beta$ from $0$ to $\infty$. 
This plot is indicated in \fig{rate_dist_gauss} as Top-$K$ ternary. 

\comment{
The particular shape of the plot can be explained as follows. 
Out of the two ends of the line plot, $\beta=0$ corresponds to the end that coincides with Scaled-sign, and $\beta=\infty$ corresponds to end that goes to $(1,0)$ coordinate. 
We observe that as $\beta$ increases, rate increases first and then decreases. 
This is due to the use of a ternary alphabet, which yields the maximum entropy when the three quantization levels are equiprobable. 
This event corresponds to the $\beta$ that makes $p_0=p_1$, $\beta=0.431$. 
A similar reasoning applies to how $D/\sigma^2$ first decreases and then increases. 
One can show that the minimum distortion is yielded at a $\beta$ that makes all three quantizing regions incur the same distortion. 
We note that all rate distortion pairs to the right and above of the Top-$K$ 1-bit plot is achievable through this proposed scheme. Therefore, we can disregard the impact of the ring-like shape in the plot.}

\section{Vector quantizers} \label{secn:ecvq}
A natural generalization of entropy-constrained scalar quantization is \emph{entropy-constrained vector quantization}. 
The latter solves for the optimal quantizing regions in high dimensional space, and the corresponding reconstruction points. 
In this section we redefine $X$ and $\hX$ to be a $N$ dimensional vectors. 
The entries of $X\in\Rel^N$ are normally distributed \iid{} random variables with variance $\sigma$. 
Let $\cc{R}_0,\dots,\cc{R}_{M-1}$ be $M$ quantizing regions in the $N$ dimensional space. 
The reconstruction of $X$ is $\hX\in\Rel^N$, and $\hX$ is assigned one of $s_0,\dots,s_{M-1}$ reconstruction points that are in the $N$ dimensional space. 
Let $p_0,\dots,p_{M-1}$ be the probabilities each of the $M$ reconstruction points are picked. 
The entropy-constrained vector quantization problem can be expressed as 
\begin{align}
\min_{\begin{subarray}{l}
	s_0,\dots,s_{M-1}\\
	\cc{R}_0,\dots,\cc{R}_{M-1}
	\end{subarray}} \quad & \EV[\Ltwo{X-\hX}^2]\\
\textrm{s.t.} \quad\quad & H(\hX) \leq R, 
\end{align}
where $H(\hX) = - \sum_{m=0}^{M-1} p_m\log p_m$. 
Similar to \eqn{lagrangian} for the scalar case, we can obtain an unconstrained optimization problem by writing the Lagrangian with parameter $\lambda\geq0$. 
The solution to the Lagrangian minimization is obtained by applying the Chou-Lookabaugh-Gray (CLG) algorithm \cite{chou1989entropy}, the extension of the Lloyd-Max algorithm to vector quantization.

\subsection{Chou-Lookabaugh-Gray algorithm} 
It may not be straightforward to describe the exact boundaries of the optimal quantizing regions $\cc{R}_0,\dots,\cc{R}_{M-1}$. 
Therefore, the CLG algorithm is implemented for a given dataset. 
Let $\{x_k\}$ be a set of realizations of $X$. 
Each realization $x_k$ is assigned to one reconstruction point out of $M$ possibilities $s_0,\dots,s_{M-1}$. 
Let $p_0,\dots,p_{M-1}$ be the fractions of data points that corresponds to the $M$ reconstruction points. 
The CLG algorithm is initialized by setting a set of initial guesses to $s_0,\dots,s_{M-1}$, and then the algorithm is iterated executing the following two steps: 
\begin{align}
B_m &= \{ x_k \mid m = \textstyle\argmin_{m'} \Ltwo{x_k-s_{m'}}^2 - \lambda\log p_{m'} \}
\quad \text{ for } m=1,\dots,M-1 \label{eqn:clgstepone} \\
s_m &= \frac{1}{|B_m|} \sum_{x_k\in B_m} x_k
\qquad\qquad\qquad\qquad\qquad\qquad\qquad\, \text{ for } m=0,\dots,M-1. 
\end{align}
Here, $|B_m|$ denotes the cardinality of $B_m$. 
The first step computes the set of data points associated with the $m$th reconstruction region, 
and the second step computes the optimal reconstruction point that minimizes the distortion. 
This iteration is continued until the distortion averaged across the data points converges. 
For different $\lambda\geq0$ we obtain operating points of the quantizer corresponding to different entropy rates. 
Note that setting $\lambda=0$ recovers the $K$-means clustering algorithm. 
A zero $\lambda$ means there is no constraint on the rate. 
When $\lambda=0$, $x_k$ is assigned to the closet reconstruction point. 
The impact of having a non-zero $\lambda$ can be easily understood by assuming a large $\lambda$. 
In that case, the minimization in \eqn{clgstepone} favors a reconstruction point $s_{m'}$ corresponding to a higher $p_{m'}$, even though $s_{m'}$ may not be the closest to $x_k$. 
This assignment creates an asymmetry in the probability distribution $p_{m'}$, making the entropy smaller. 
Intuitively, having one or two reconstruction points with very large $p_{m'}$ incurs a smaller entropy than having a large number of reconstruction points with equal $p_{m'}$. 

\def\threewidth{.32}

\subsection{Visualizing optimal quantization regions} 
\begin{figure}
\centering\includegraphics[width=\threewidth\textwidth]{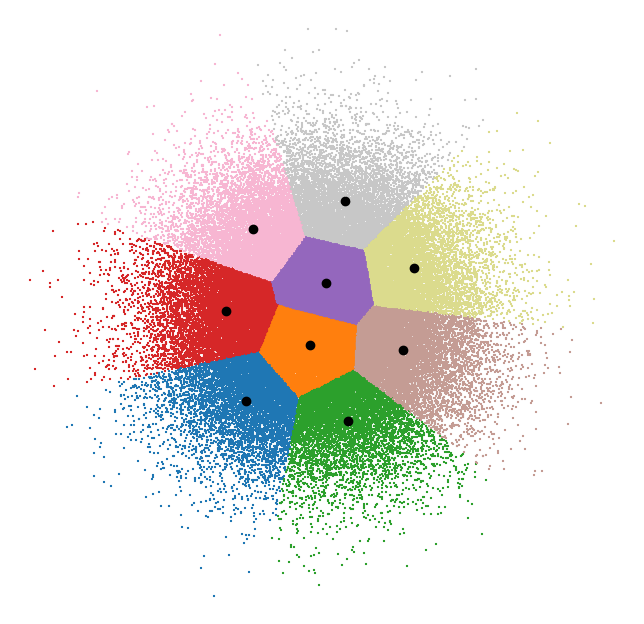}
\centering\includegraphics[width=\threewidth\textwidth]{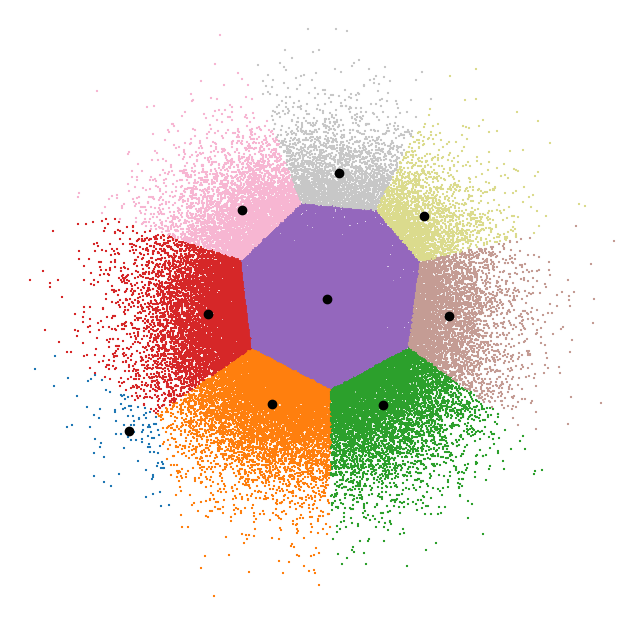}
\centering\includegraphics[width=\threewidth\textwidth]{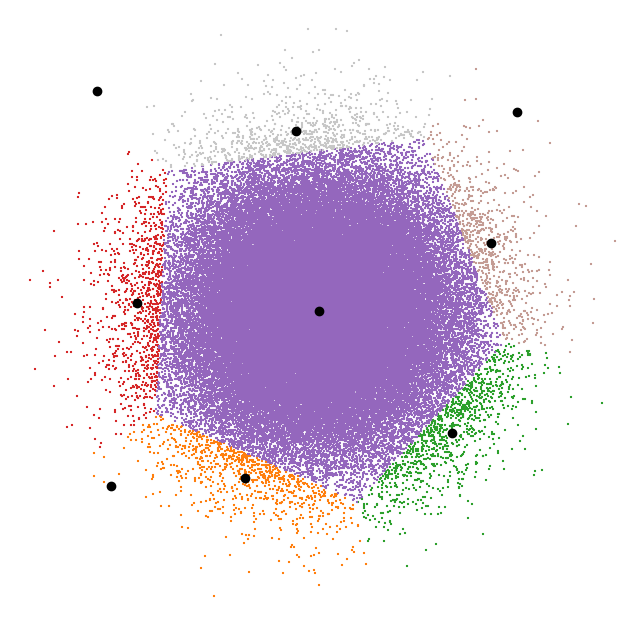}
\caption{Optimal assignment of $10^5$ normally distributed $2$-d random vectors to $9$ reconstruction points. From left to right sub-figures correspond to three values of $\lambda$; $0$, $0.4$ and $0.8$.} 
\label{fig:clgwithlambda}
\end{figure}

\fig{clgwithlambda} illustrates the how $B_m$ sets are determined by CLG for $N=2$ and $M=9$. 
Setting $\lambda=0$ gives the left-most sub-figure in which the algorithm finds $B_m$ and $s_m$ that minimize distortion, without any constraint on entropy. 
This yields a uniform distribution across the reconstruction points with $p_m\approx1/9$ producing the maximum entropy for $M=9$. 
As $\lambda$ gets larger, such as in the middle and right-most sub-figures, most of the probability mass is assigned to one region. 
For example, in the right-most sub-figure 95682 $x_k$ are assigned to the reconstruction point in middle. 
The five reconstruction points immediately around the one in middle get roughly equal number of assignments: 713, 840, 873, 937, and 955 $x_k$. 
The three outermost reconstruction points are not assigned any $x_k$. 
For $N=2$ and $\lambda=0.8$, even if we set $M$ larger than $9$, the CLG algorithm finds $6$ quantizing regions with non-zero probabilities that are similar to those in the right-most sub-figure.

\subsection{Comparing with scalar quantization}
\begin{figure}
\centering\includegraphics[width=\threewidth\textwidth]{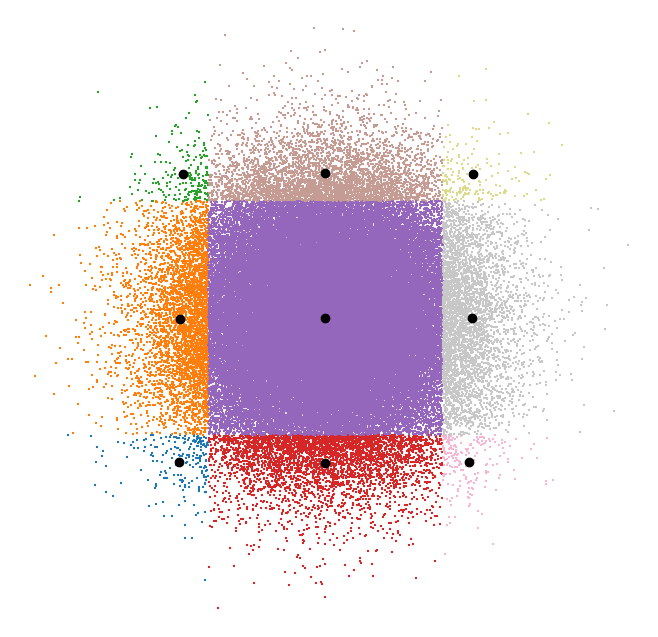}
\centering\includegraphics[width=\threewidth\textwidth]{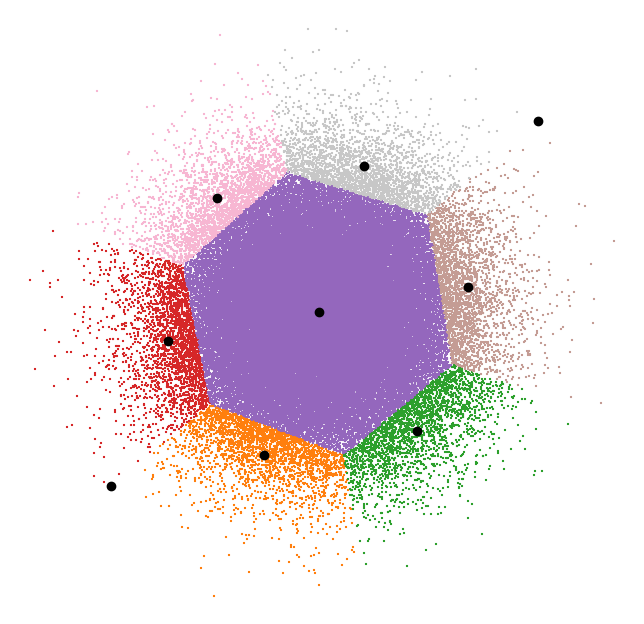}
\caption{Quantization of $10^5$ normally distributed $2$-d random vectors. Left and right sub-figures correspond to Lloyd-Max $M=3$ ($3$ reconstruction points per random variable, $9$ points for two) and CLG $M=9$. 
Parameters in each method is tuned to produce a distortion of $0.6$.} 
\label{fig:comparesqvsvq}
\end{figure}

For rates below one bit, as per \fig{rate_dist_gauss}, Lloyd-Max $M=3$ (Top-$K$ ternary) quantizer performs as well as any other scalar quantizer with a larger $M$. 
For this reason we choose Lloyd-Max $M=3$ and illustrate its quatization regions in 2-d space, i.e., when two random variables are quantized using $9$ reconstruction points. 
We compare these quatization regions with the ones produced by CLG when $N=2$ and $M=9$. 
The two sub-figures in \fig{comparesqvsvq} presents the comparison. 
Note that Lloyd-Max $M=3$ quntizer is only able to produce rectangular boundaries in 2-d space since the two dimensions are considered independently. 
In comparison, CLG produces more complex boundaries by jointly considering the two dimensions. 
The two outermost reconstruction points in CLG not not assigned any $x_k$. 
Both quantizers produce a distortion of $0.6$. 
The corresponding entropy of Lloyd-Max and CLG quantizers are $0.524$ and $0.511$ respectively. 
Therefore, CLG achieves a lower entropy for the same distortion by jointly quantizing two random variables. 

We can numerically compute entropy $H(\hX)$ and distortion $\EV[\Ltwo{X-\hX}^2]$ of the CLG quantizer trained on $\{x_k\}$. 
We compute and plot in \fig{rate_dist_clg} such rate distortion pairs for $N=2,\dots,8$. 
Different rates are obtained by varying $\lambda$. 
Recall that CLG algorithm requires an $M$ parameter that indicates the number of reconstruction points in the quantizer. 
For each $N$ we increase $M$ given to CLG such that the algorithm assigns $p_m=0$ for at least one reconstruction point. 



\begin{thebibliography}{10}
\providecommand{\url}[1]{#1}
\csname url@samestyle\endcsname
\providecommand{\newblock}{\relax}
\providecommand{\bibinfo}[2]{#2}
\providecommand{\BIBentrySTDinterwordspacing}{\spaceskip=0pt\relax}
\providecommand{\BIBentryALTinterwordstretchfactor}{4}
\providecommand{\BIBentryALTinterwordspacing}{\spaceskip=\fontdimen2\font plus
\BIBentryALTinterwordstretchfactor\fontdimen3\font minus
  \fontdimen4\font\relax}
\providecommand{\BIBforeignlanguage}[2]{{%
\expandafter\ifx\csname l@#1\endcsname\relax
\typeout{** WARNING: IEEEtran.bst: No hyphenation pattern has been}%
\typeout{** loaded for the language `#1'. Using the pattern for}%
\typeout{** the default language instead.}%
\else
\language=\csname l@#1\endcsname
\fi
#2}}
\providecommand{\BIBdecl}{\relax}
\BIBdecl

\bibitem{gao2019rate}
W.~Gao, Y.-H. Liu, C.~Wang, and S.~Oh, ``Rate distortion for model compression:
  From theory to practice,'' in \emph{Int. Conf. Machine Learning}, California,
  Jan 2019.

\bibitem{glorot2010understanding}
X.~Glorot and Y.~Bengio, ``Understanding the difficulty of training deep
  feedforward neural networks,'' in \emph{Int. Conf. Artificial Intelligence
  and Stat.}, Sardinia, Italy, Mar 2010.

\bibitem{bernstein2018signsgd}
J.~Bernstein, Y.-X. Wang, K.~Azizzadenesheli, and A.~Anandkumar, ``{signSGD}:
  Compressed optimisation for non-convex problems,'' in \emph{Int. Conf.
  Machine Learning}, Stockholm, Jul 2018.

\bibitem{shi2019understanding}
S.~Shi, X.~Chu, K.~C. Cheung, and S.~See, ``Understanding {Top}-k
  sparsification in distributed deep learning,'' \emph{arXiv preprint
  arXiv:1911.08772}, Nov 2019.

\bibitem{cover1999elements}
T.~M. Cover, \emph{Elements of information theory}.\hskip 1em plus 0.5em minus
  0.4em\relax John Wiley \& Sons, 1999.

\bibitem{stich2018sparsified}
S.~U. Stich, J.-B. Cordonnier, and M.~Jaggi, ``Sparsified {SGD} with memory,''
  in \emph{Advances in Neural Inf. Proc. Sys.}, Montréal, Dec 2018, pp.
  4447--4458.

\bibitem{karimireddy2019error}
S.~P. Karimireddy, Q.~Rebjock, S.~U. Stich, and M.~Jaggi, ``Error feedback
  fixes {SignSGD} and other gradient compression schemes,'' \emph{Int. Conf.
  Machine Learning}, Jan 2019.

\bibitem{dryden2016communication}
N.~Dryden, T.~Moon, S.~A. Jacobs, and B.~Van~Essen, ``Communication
  quantization for data-parallel training of deep neural networks,'' in
  \emph{Workshop on Machine Learning in HPC Environments}.\hskip 1em plus 0.5em
  minus 0.4em\relax IEEE, 2016.

\bibitem{wiegand2011source}
T.~Wiegand and H.~Schwarz, \emph{Source coding: Part {I} of fundamentals of
  source and video coding}.\hskip 1em plus 0.5em minus 0.4em\relax Now
  Publishers Inc, 2011.

\bibitem{lloyd1982least}
S.~Lloyd, ``Least squares quantization in {PCM},'' \emph{IEEE Trans.\ Inform.\
  Theory}, pp. 129--137, 1982.

\bibitem{seide20141}
F.~Seide, H.~Fu, J.~Droppo, G.~Li, and D.~Yu, ``1-bit stochastic gradient
  descent and application to data-parallel distributed training of speech
  {DNNs},'' in \emph{Conf. Int. Speech Comm. Association}, Singapore, Sep 2014.

\bibitem{strom2015scalable}
N.~Strom, ``Scalable distributed {DNN} training using commodity {GPU} cloud
  computing,'' in \emph{Conf. Int. Speech Comm. Association}, 2015.

\bibitem{aji2017sparse}
A.~F. Aji and K.~Heafield, ``Sparse communication for distributed gradient
  descent,'' \emph{arXiv preprint arXiv:1704.05021}, 2017.

\bibitem{lin2017deep}
Y.~Lin, S.~Han, H.~Mao, Y.~Wang, and W.~J. Dally, ``Deep gradient compression:
  Reducing the communication bandwidth for distributed training,'' \emph{arXiv
  preprint arXiv:1712.01887}, 2017.

\bibitem{wangni2018gradient}
J.~Wangni, J.~Wang, J.~Liu, and T.~Zhang, ``Gradient sparsification for
  communication-efficient distributed optimization,'' in \emph{Advances in
  Neural Inf. Proc. Sys.}, Montréal, Dec 2018, pp. 1299--1309.

\bibitem{chou1989entropy}
P.~A. Chou, T.~Lookabaugh, and R.~M. Gray, ``Entropy-constrained vector
  quantization,'' \emph{IEEE Trans.\ Acoust., Speech, Signal Processing}, pp.
  31--42, 1989.

\end{thebibliography}
\end{document}